\documentstyle[psfig]{mn}

\title[Spiral galaxy distance indicators based on near-infrared 
       photometry]
      {Spiral galaxy distance indicators based on near-infrared 
       photometry\thanks{Based on observations obtained at the 
       European Southern Observatory, La Silla, Chile}}

\author[R.  de Grijs and R.F.  Peletier]{Richard de
Grijs$^{1,2}$\thanks{E-mail: grijs@virginia.edu} and Reynier F. 
Peletier$^3$\thanks{Present address: School of Physics and Astronomy,
University of Nottingham, University Park, Nottingham NG7 2RD} \\
$^1$ Astronomy Department, University of Virginia, PO Box 3818,
Charlottesville, VA 22903-0818, USA \\
$^2$ Kapteyn Astronomical Institute, University of Groningen, PO Box
800, 9700 AV Groningen, the Netherlands \\
$^3$ Department of Physics, University of Durham, South Road, Durham
DH1 3LE, United Kingdom}

\date{Received date; accepted date}
\pubyear{1999}

\begin{document}
\maketitle

\begin{abstract}
We compare two methods of distance determination to spiral galaxies
using optical/near-infrared (NIR) observations, the {\it (I--K)} versus
$M_K$ colour -- absolute magnitude (CM) relation and the {\it I} and
{\it K}-band Tully-Fisher relation (TFR).  \\
Dust-free colours and NIR absolute magnitudes greatly enhance the
usefulness of the NIR CM relation as a distance indicator for moderately
to highly inclined {\em spiral} galaxies {\it in the field}
(inclinations between $\sim 80^\circ$ and $90^\circ$); by avoiding
contamination by dust the scatter in the CM relation is significantly
reduced, compared to similar galaxy samples published previously.  The
CM relation can be used to determine distances to field spiral galaxies
with $M_K > -25.5$, to at least $M_K \approx -20$.  \\
Our results, supplemented with previously published observations for
which we can -- to some degree -- control the effects of extinction, are
consistent with a universal nature of the CM relation for field spiral
galaxies.  \\
High-resolution observations done with the {\sl Hubble Space Telescope}
can provide a powerful tool to calibrate the relation and extend the
useful distance range by more than a factor of 2 compared to
ground-based observations. \\
The intrinsic scatter in the NIR CM relation in the absolute {\it
K}-band magnitudes is $\sim 0.5$ mag, yielding a lower limit to the
accuracy of distance determinations on the order of 25\%.  \\
Although we find an unusually low scatter in the TFR (probably a
statistical accident), a typical scatter in the TFR would yield
distances to our sample galaxies with uncertainties of only $\sim 15$\%. 
However, one of the main advantages of the use of the NIR CM relation is
that {\em we only need photometric data} to obtain distance estimates;
use of the TFR requires additional kinematic data, although it can be
used to significantly greater distances. 
\end{abstract}

\begin{keywords}
distance scale -- galaxies: photometry -- galaxies: spiral -- galaxies:
statistics -- infrared: galaxies
\end{keywords}

\section{Photometric distance indicators}

The study of highly-inclined disc-dominated galaxies offers the unique
opportunity to avoid the confusing effects of interstellar extinction on
the optical and near-infrared (NIR) appearance of their dominant (old)
stellar disc population.  

In this paper we use NIR absolute magnitudes and {\it I--K} colours that
were determined in those regions of our statistically complete sample of
edge-on field galaxies that are least affected by extinction, to derive
a ``dust-free'' colour -- absolute magnitude (CM) relation.  Dust-free
colours and NIR absolute magnitudes greatly enhance the usefulness of
the NIR CM relation as a distance indicator for moderately to highly
inclined spiral galaxies in the field. 

Alternatively, large galaxy surveys in the NIR facilitate the use of the
NIR Tully-Fisher relation (TFR; Tully \& Fisher 1977) as an accurate
tool to obtain distances to spiral galaxies in clusters.  In this paper
we discuss the {\it I} and {\it K}-band TFRs derived from our sample
galaxies and the accuracy with which they can be used to estimate
distances to field spiral galaxies. 

\section{The NIR colour-magnitude relation}

The tightness of the CM relation for {\it early-type} galaxies (as first
established by Baum [1959] and de Vaucouleurs [1961]), makes it
potentially useful as a distance indicator, as was first suggested by
Sandage (1972).  In this paper we investigate whether the NIR
CM relation is also useful as a diagnostic tool to estimate distances to
{\it spiral galaxies in the field}. 

\subsection{Elliptical versus spiral galaxies}
\label{EvsS.sect}

Visvanathan \& Griersmith (1977) extended the range of galaxy types from
elliptical/S0 galaxies to early-type spirals (S0/a to Sab), and found
(within the errors) exactly the same optical CM relation for the
early-type Virgo cluster spirals as had been found for E/S0 galaxies,
but with a larger excess {\it scatter} (see also Visvanathan \& Sandage
1977; Griersmith 1980; Bower, Lucey \& Ellis 1992b; Peletier \& de Grijs
1998). 

Despite the good agreement between the elliptical and spiral galaxy CM
relations, later-type spiral galaxies occupy a different region in the
CM diagram (e.g., Tully, Mould \& Aaronson 1982; Mobasher, Ellis \&
Sharples 1986; Peletier \& de Grijs 1998).  Apart from this distinction
between the loci of late-type spirals and E/S0's, Griersmith (1980) also
noticed that differences in zero point of the CM relations seem to
follow a systematic trend along the Hubble sequence: colours become
systematically bluer for later Hubble types (see also Peletier \& de
Grijs 1998), although the slopes for the individual galaxy types are the
same (within the errors) compared to each other and to E/S0 galaxies. 

\subsection{A universal CM relation?}
\label{universal.sect}

Following the discovery of the CM effect for the early-type Virgo
cluster members, many attempts have been made to unambiguously determine
the universality of the relationship (e.g., Visvanathan \& Sandage 1977;
Visvanathan \& Griersmith 1977; Griersmith 1980; Aaronson, Persson \&
Frogel 1981; Bower et al.  1992a,b).  Provided that the CM relation is
universally applicable it can be used as a distance indicator.  In
general, the UV-optical CM relation {\it for early-type galaxies} was
found to be universal and independent of environment to a high degree
(for a recent review see Ellis et al.  1997).  However, the tests done
to study the universality of the CM effect were challenged by a number
of studies (e.g., Faber 1977; Burstein 1977; Larson, Tinsley \& Caldwell
1980; Aaronson et al.  1981, and more recently Abraham et al.  1996 and
van Dokkum et al.  1998), claiming that non-negligible environmental
effects were playing a role in the observational evidence.  To undertake
an independent study of the universality of the CM relation, Bower et
al.  (1992a,b) obtained new observations of Virgo and Coma cluster
galaxies.  Their observations support the argument that a universal CM
relation for cluster E/S0 galaxies, over the entire wavelength range, is
likely, although the dispersion is expected to be considerable.  Bower
et al.'s (1992b) photometry allows the CM relation for early-type
cluster galaxies to be used to estimate distances accurate to $\sim$20\%
per galaxy. 

In this paper we will use our own optical and NIR observations of
edge-on spiral galaxies, supplemented with similar samples from the
literature, to investigate the possible universality of the CM relation
for {\it spiral galaxies in the field}. 

\subsection{Advantages of NIR observations}

For the E/S0 galaxies in the Virgo cluster the change of colour with
absolute magnitude is greatest in the ultraviolet, and decreases
significantly towards redder wavelengths (Visvanathan \& Sandage 1977,
and references therein).  However, the CM effect shows up again in the
optical-NIR regime, in the sense that {\it V--K} is bluer for
intrinsically fainter galaxies (e.g., Aaronson et al.  1981, Tully et
al.  1982).  The usefulness of NIR observations of spiral galaxies in
the field for measuring extragalactic distances depends on two main
questions (e.g., Aaronson et al.  1981):

\begin{enumerate} 
\item Do {\it spiral} galaxies in the field follow a universal CM
relation in the NIR;
\item If so, is the scatter sufficiently small for a useful application
of the relation?
\end{enumerate}

The main advantages of NIR observations compared to observations in the
blue passbands are their relative insensitivity to contamination by the
presence of young stellar populations and dust.  The absorption
corrections for dust in external galaxies, which are largest in the
blue, are difficult and controversial; therefore, we were motivated to
study the CM relation for our sample of edge-on spiral galaxies in the
NIR, using {\it I--K} colours and absolute {\it K}-band magnitudes. 

In Peletier \& de Grijs (1998) we find that the scatter in the NIR CM
relation for (field) spiral galaxies can entirely be explained by
observational uncertainties.  Moreover, we show that the slope of the
relation is steeper for spirals than for ellipticals.  We can explain
this if we assume that the CM relations for spiral and elliptical
galaxies are intrinsically different, in the sense that the stars in
spiral galaxies are younger than those in ellipticals and the fraction
of young stars in a spiral galaxy (i.e., its ``age'') is determined
solely by the galaxy's luminosity, and not by its environment.  The CM
relation formed by elliptical galaxies, on the other hand, is generally
attributed to changes in metallicity (Peletier \& de Grijs 1998, and
references therein). 

In Sect.  \ref{dustfree.sect} we discuss the nature of our observations
and the corrections that have to be applied for an accurate study of the
NIR CM relation.  We supplement our data with recently published {\it I}
and {\it K}-band data in Sect.  \ref{composite.sect}.  In Sect.  5 we
discuss the implications of our results and look into the applicability
of a tight NIR CM relation using {\sl HST} observations.  We discuss the
accuracy of the NIR TFR in Sect.  6, and we give
a summary of our results and conclusions in Sect.  7. 

\section{Constructing the CM relation}
\label{dustfree.sect}

The observations of and the reduction techniques applied to our sample
of highly-inclined spiral galaxies, as well as the resulting photometric
accuracy, have been described in de Grijs (1997, 1998).  Here, we
summarize our selection criteria, applied to the galaxies in the Surface
Photometry Catalogue of the ESO-Uppsala Galaxies (ESO-LV; Lauberts \&
Valentijn 1989):
\begin{itemize} 
\item Inclination selection: $i \ge 87^\circ$; 
\item Diameter selection: $D_{25}^B \ge 2.'2$ ($D_{25}^B$: blue angular
size at a surface brightness $\Sigma_B = 25$ mag arcsec$^{-2}$); 
\item Galaxy type selection:  S0 -- Sd, and 
\item Morphology selection: they should be non-interacting and undisturbed.
\end{itemize}

Although our NIR observations are relatively insensitive to disturbing
dust effects (see, e.g., de Grijs, Peletier \& van der Kruit 1997), we
need to apply several corrections to the observed quantities before they
can be used to study the CM relation. 

\subsection{Colour gradients and Galactic foreground extinction}

Colour gradients in edge-on galaxies have not been studied extensively. 
In a few well-studied edge-on spiral galaxies, colour gradients parallel
to the major axis were found to be negligible (e.g., Hamabe et al. 
1980, Jensen \& Thuan 1982, van der Kruit \& Searle 1982), or show an
increasingly blue disc population with increasing radius (e.g., Sasaki
1987, Aoki et al.  1991), although such gradients are generally small. 

Vertical colour gradients in the discs of our sample edge-on galaxies are
generally small or negligible (de Grijs et al.  1997, de Grijs \&
Peletier, in preparation); in fact, the vertical colour gradients are
smaller than or of the same order as the observational errors for our
{\it I--K} colour determinations.  The only statistically significant
vertical colour variation is the reddening due to dust in the
galactic planes.  As we showed in de Grijs \& van der Kruit (1996) and
de Grijs \& Peletier (1997), the vertical scale height does not vary
significantly as a function of wavelength, which indirectly shows the
absence of any significant vertical colour gradient. 

In de Grijs et al.  (1997) we show that, once away from the dust lane,
the vertical {\it I--K} colour profiles are generally (approximately)
flat and featureless, indicating that the excess extinction in the {\it
I} band (compared to that in the {\it K} band) is negligible at these
heights above the galactic planes.  The colours we use in this paper
were determined from this (approximately) flat part of the vertical
colour profile at the minor axis (at $1.5 h_z \le |z| \le 3.5 h_z$,
where $h_z$ is the exponential scale height), to avoid the reddening
caused by the in-plane dust.  Since young stellar populations are
generally confined to regions close to the galactic planes, an
additional advantage of determining colours away from the planes is that
the contamination by emission from these young stellar populations is
thus greatly reduced or negligible. 

Peletier \& Balcells (1997) showed, for their sample of 30 field spiral
and lenticular galaxies, that the bulge colours on the minor axis and
the inner disc colours taken in wedge apertures at 15$^\circ$ from the
major axis at 2 {\it K}-band scale lengths are very similar.  Therefore,
the colour of an edge-on disc-dominated galaxy determined in a dust-free
region at the minor axis can be considered as representative for the
galaxy's dominant (old) stellar population, under the assumption that
vertical colour gradients in the old-disc population are smaller than
the observational uncertainties. 

To deal with Galactic reddening, we used the Galactic extinction values
in the {\it I} and $K'$ bands, $A_{G,I}$ and $A_{G,K'}$ given by
Schlegel, Finkbeiner \& Davis (1998), thereby assuming that the Galactic
extinction can be approximated by a foreground dust screen.  These
full-sky dust maps are twice as accurate as the older reddening
estimated by Burstein \& Heiles (1978, 1984) in regions of low and
moderate reddening and likely significantly more accurate in regions of
high reddening (see also Hudson 1999).  The Galactic extinction in the
{\it I} and $K'$ bandpasses was obtained assuming a standard $R_V = 3.1$
extinction law (Schlegel et al.  1998, their Appendix B). 

\subsection{Cosmological corrections and distance calibration}

The corrections for the effects of redshift ({\it K}-corrections) are
generally small in the NIR.  For our sample galaxies the {\it
K}-corrections for the {\it I--K} colours range from 0.00 -- 0.04 mag
(e.g., Schneider, Gunn \& Hoessel 1983), depending on galaxy type (e.g.,
Coleman, Wu \& Weedman 1980), see Table \ref{corrections.tab}. 

Heliocentric velocities for the majority of our sample galaxies were
obtained by Mathewson, Ford \& Buchhorn (1992) and Mathewson \& Ford
(1996), which provides us with a homogeneous data set to base our
absolute magnitude calculations on.  To obtain absolute magnitudes we
applied the formula for the systemic velocities adjusted for the solar
motion with respect to the centroid of the Local Group given by Richter,
Tammann \& Huchtmeier (1987, see also Schmidt \& Boller 1992):
\begin{equation} 
v_{LG} = v_\odot + \Delta v \; ,
\end{equation} 
where
\begin{eqnarray} 
\Delta v = -49.59 \cos \ell \cos b &+& 306.95 \sin \ell \cos b \nonumber \\
&-& 18.59 \sin b
\end{eqnarray}

In Table \ref{corrections.tab} we give an overview of the observed and
derived quantities used for the study of both the NIR CM and TF
relations. 

{
\begin{table*}

\caption[]{\label{corrections.tab}{\bf Basic properties of the sample
galaxies}\\
Columns: (1) Galaxy name (ESO-LV); (2) heliocentric velocity (Mathewson
et al.  1992); (3) velocity correction for motion w.r.t.  the centroid
of the Local Group (Richter et al.  1987); (4)--(5) and (6)--(7)
apparent {\it I} and {\it K}-band magnitudes and observational errors;
(8) Galactic extinction in $K'$; (9) {\it I--K} colour excess (Schlegel
et al.  1998); (10) and (11) {\it I--K} color, corrected for Galactic
foregound extinction, and observational error; (12) {\it K}-correction
(interpolated from Schneider et al.  1980); (13) absolute {\it K}-band
magnitude ($H_0 = 100$ km s$^{-1}$ Mpc$^{-1}$), corrected for Galactic
foreground extinction and the effects of redshift; (14) peculiar
velocity with respect to the centroid of the Local Group (Mathewson et
al. 1992).}

\begin{center}
\tabcolsep=1mm

\begin{tabular}{crrrcrcccccccr}
\hline
Galaxy & \multicolumn{1}{c}{$v_\odot$} & \multicolumn{1}{c}{$\Delta v$}
& $m_I^0$ & $\pm$ & $m_K^0$ & $\pm$ & $A_{G,K'}$ & $E(I-K)$ & $(I-K)_0$
& $\pm$ & {\it K}-corr.  & $M_K^0$ & $v_{\rm pec,LG}$ \\
\noalign{\vspace{2pt}}
\cline{4-13}
\noalign{\vspace{1pt}}
(ESO) & (km/s) & (km/s) & \multicolumn{10}{c}{(mag)} & (km/s) \\
(1) & \multicolumn{1}{c}{(2)} & \multicolumn{1}{c}{(3)} & \multicolumn{1}{c}{(4)} & (5) & 
\multicolumn{1}{c}{(6)} & (7) & (8) & (9) & (10) & (11) & (12) & (13) & 
\multicolumn{1}{c}{(14)} \\
\\
026 -G 06 & 2748 & -202.1 & 12.97 & 0.02 & 11.45 & 0.09 & 0.05 & 0.22 & 1.21 & 0.10 & 0.01 & -20.38 &    49 \\
141 -G 27 & 1922 & -142.3 & 12.59 & 0.02 & 10.55 & 0.12 & 0.02 & 0.10 & 1.41 & 0.05 & 0.01 & -20.26 &   150 \\
142 -G 24 & 2119 & -125.4 & 12.14 & 0.01 & 10.53 & 0.16 & 0.03 & 0.13 & 1.19 & 0.20 & 0.01 & -21.01 &  -344 \\
157 -G 18 & 1268 & -206.9 & 12.26 & 0.02 &  9.78 & 0.17 & 0.01 & 0.03 & 1.33 & 0.10 & 0.00 & -20.86 &  -304 \\
201 -G 22 & 4014 & -183.9 & 13.06 & 0.03 & 10.33 & 0.16 & 0.01 & 0.03 & 1.62 & 0.30 & 0.01 & -22.90 &  -835 \\
263 -G 15 & 2525 & -310.5 & 10.82 & 0.01 &  9.02 & 0.15 & 0.07 & 0.30 & 1.69 & 0.10 & 0.01 & -22.70 &    -- \\
286 -G 18 & 9162 &  -34.9 & 12.12 & 0.01 & 10.42 & 0.11 & 0.02 & 0.06 & 1.71 & 0.07 & 0.03 & -24.44 &  -554 \\
311 -G 12 & 1128 & -279.7 &  9.29 & 0.01 &  7.21 & 0.07 & 0.14 & 0.61 & 1.55 & 0.05 & 0.00 & -22.44 &    -- \\
315 -G 20 & 4843 & -300.3 & 12.19 & 0.01 &  9.11 & 0.19 & 0.08 & 0.35 & 1.40 & 0.07 & 0.02 & -24.18 &    -- \\
340 -G 09 & 2546 &  -20.1 & 13.51 & 0.07 & 11.32 & 0.16 & 0.02 & 0.09 & 1.20 & 0.15 & 0.01 & -20.62 &   -58 \\
358 -G 29 & 1776 & -119.1 & 10.34 & 0.01 &  8.25 & 0.09 & 0.01 & 0.02 & 1.43 & 0.05 & 0.01 & -22.85 &    -- \\
383 -G 05 & 3637 & -240.1 & 11.83 & 0.02 &  8.93 & 0.22 & 0.02 & 0.09 & 1.61 & 0.03 & 0.01 & -23.73 &    -- \\
416 -G 25 & 4998 &  -71.6 & 12.46 & 0.03 & 10.34 & 0.18 & 0.01 & 0.05 & 1.56 & 0.08 & 0.02 & -23.45 & -1118 \\
435 -G 14 & 2697 & -285.9 & 12.38 & 0.05 & 10.09 & 0.10 & 0.03 & 0.15 & 1.65 & 0.12 & 0.01 & -22.76 & -1522 \\
435 -G 25 & 2470 & -288.3 & 10.89 & 0.03 &  8.39 & 0.11 & 0.03 & 0.13 & 1.79 & 0.10 & 0.01 & -23.49 &  -262 \\
437 -G 62 & 2850 & -289.8 & 10.57 & 0.01 &  8.31 & 0.19 & 0.03 & 0.12 & 1.71 & 0.20 & 0.01 & -23.73 &    -- \\
446 -G 18 & 4843 & -204.9 & 12.59 & 0.02 & 10.15 & 0.10 & 0.02 & 0.10 & 1.71 & 0.25 & 0.02 & -23.33 &  -660 \\
446 -G 44 & 2793 & -204.2 & 12.37 & 0.04 & 10.20 & 0.12 & 0.03 & 0.11 & 1.58 & 0.05 & 0.01 & -22.35 &  -847 \\
460 -G 31 & 5759 &   25.3 & 12.07 & 0.02 &  9.94 & 0.15 & 0.08 & 0.33 & 1.82 & 0.15 & 0.02 & -23.70 &   216 \\
487 -G 02 & 1755 & -153.8 & 11.38 & 0.01 &  8.83 & 0.12 & 0.01 & 0.04 & 1.57 & 0.05 & 0.01 & -22.77 &  -656 \\
509 -G 19 &10727 & -218.9 & 12.20 & 0.01 &  9.55 & 0.13 & 0.03 & 0.09 & 2.02 & 0.08 & 0.04 & -25.56 &    -- \\
564 -G 27 & 2178 & -259.4 & 12.12 & 0.09 &  9.66 & 0.11 & 0.06 & 0.24 & 1.66 & 0.10 & 0.01 & -22.65 & -1166 \\
\hline
\end{tabular}
\end{center}
\end{table*}

\section{Comparison with literature samples}
\label{composite.sect}

In Fig.  \ref{observations.fig}a we have plotted the CM relation derived
from our ``dust-free'' {\it I--K} colours (determined in the region $1.5
h_z \le |z| \le 3.5 h_z$) and the absolute {\it K}-band magnitudes,
$M_K^0$, both corrected for Galactic foreground extinction.  To assess
the importance of contamination by dust, in Fig. 
\ref{observations.fig}b we compare the NIR CM relation resulting from
the use of integrated galaxy colours to that derived from the dust-free
colours.  It is clear that the importance of dust should not be
underestimated: the integrated colours are systematically redder than
the ``dust-free'' colours. 

Careful examination of Fig.  \ref{observations.fig}a reveals a rather
peculiar distribution of the absolute {\it K}-band magnitudes of our
sample galaxies, into two clumps and a single, bright galaxy.  We
believe that this is not due to any selection effects other than the
availability of observing time with NIR arrays.  The corresponding
distribution of absolute {\it I}-band magnitudes, which were obtained
for about twice as many of our sample galaxies, does not show any
significant subclumping, although the two brightest galaxies are $\sim
1$ mag brighter than the third brightest galaxy observed as part of our
project. 

\begin{figure*}
\psfig{figure=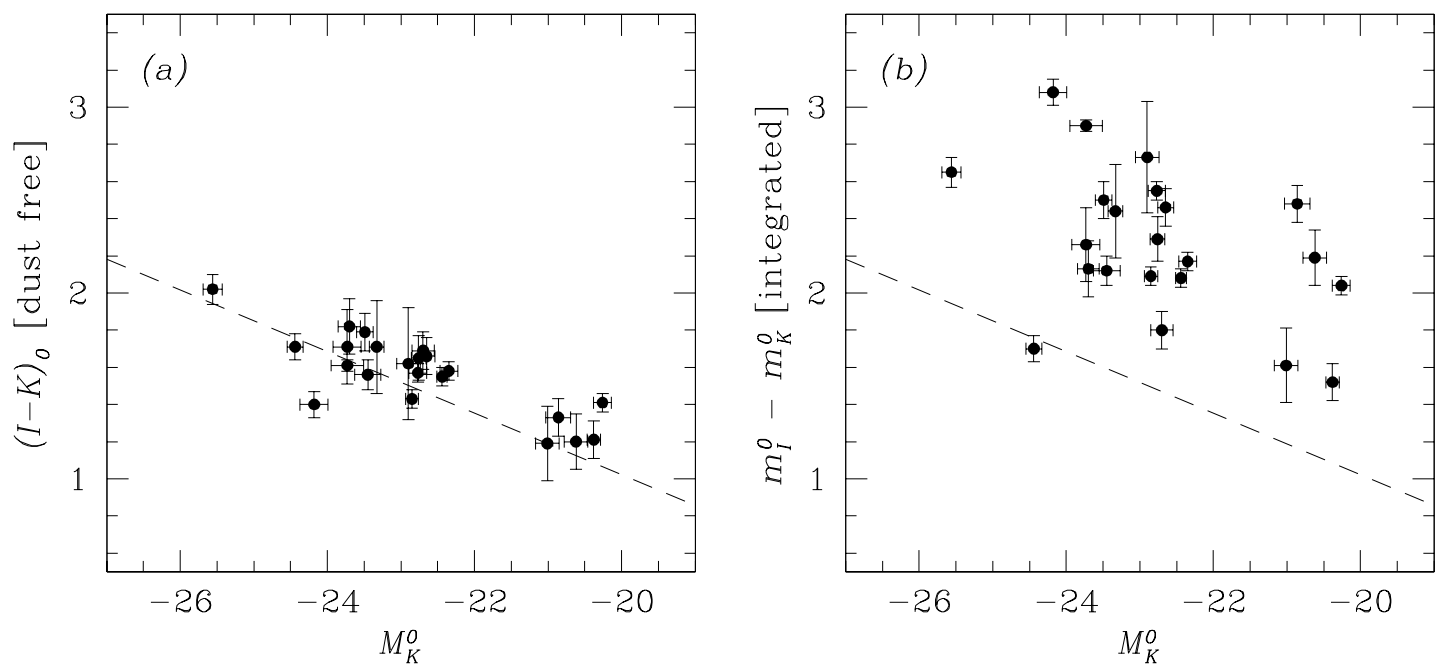}
\vspace*{-10cm}
\caption[]{\label{observations.fig}{\it (a)}
Correlation between absolute {\it K}-band magnitude, $M_K^0$, and
dust-free {\it I--K} colours.  The best-fitting NIR CM relation is
indicated by the dashed line, for which the parameters are given in Table
\ref{slopes.tab}; {\it (b)} Corresponding CM relation derived from the
integrated galaxy colours. Almost all data points are located redwards
from the best-fitting CM relation shown in Fig. \ref{observations.fig}a,
due to internal interstellar extinction. All observational quantities
have been corrected for Galactic foreground extinction.}
\end{figure*}

In Fig.  \ref{litcompare.fig} we compare our results with those of other
galaxy samples published recently.  Only very few samples with
high-quality {\it K}-band observations have been published to date:

\begin{description}

\item[{\it (a)}] The sample of Bershady et al.  (1994) and Bershady
(1995) consists of 171 field galaxies of all types, selected from three
high Galactic latitude fields (where the Galactic extinction is low), of
which a subset of 143 is statistically representative in its sampling of
the apparent colour distribution of galaxies.  The median redshift of the
sample galaxies is $z = 0.14$.  Absolute magnitudes were determined
using luminosity distances, corrected for cosmological effects. 

\item[{\it (b)}] De Jong's (1996b) sample is a diameter-limited sample
of 86 (nearly) face-on spiral galaxies.  His absolute magnitude
calculations are based on heliocentric velocities, corrected for
Virgocentric infall using the model of Kraan-Korteweg (1986; de Jong,
priv.  comm.), and taking into account the Galactic foreground
extinction estimates of Burstein \& Heiles (1978, 1984). 

\item[{\it (c)}] The data presented by Andredakis, Peletier \& Balcells
(1995) and Peletier \& Balcells (1997) consists of 37 field disc
galaxies of types S0 to Sbc, uniform in orientation on the sky, for 30
of which NIR observations are available.  To obtain absolute magnitudes,
they used the Galactic standard-of-rest corrections given by de
Vaucouleurs et al.  (1991).  Corrections for foreground extinction were
only applied to their {\it I}-band observations, since the {\it K}-band
corrections were smaller than their observational errors. 

\item[{\it (d)}] Tully et al.'s (1996) sample was taken from the Ursa
Major cluster; they present {\it I} and $K'$-band observations of a
magnitude-limited sample of 70 disc-dominated galaxies. 
\end{description}

\begin{figure*}
\psfig{figure=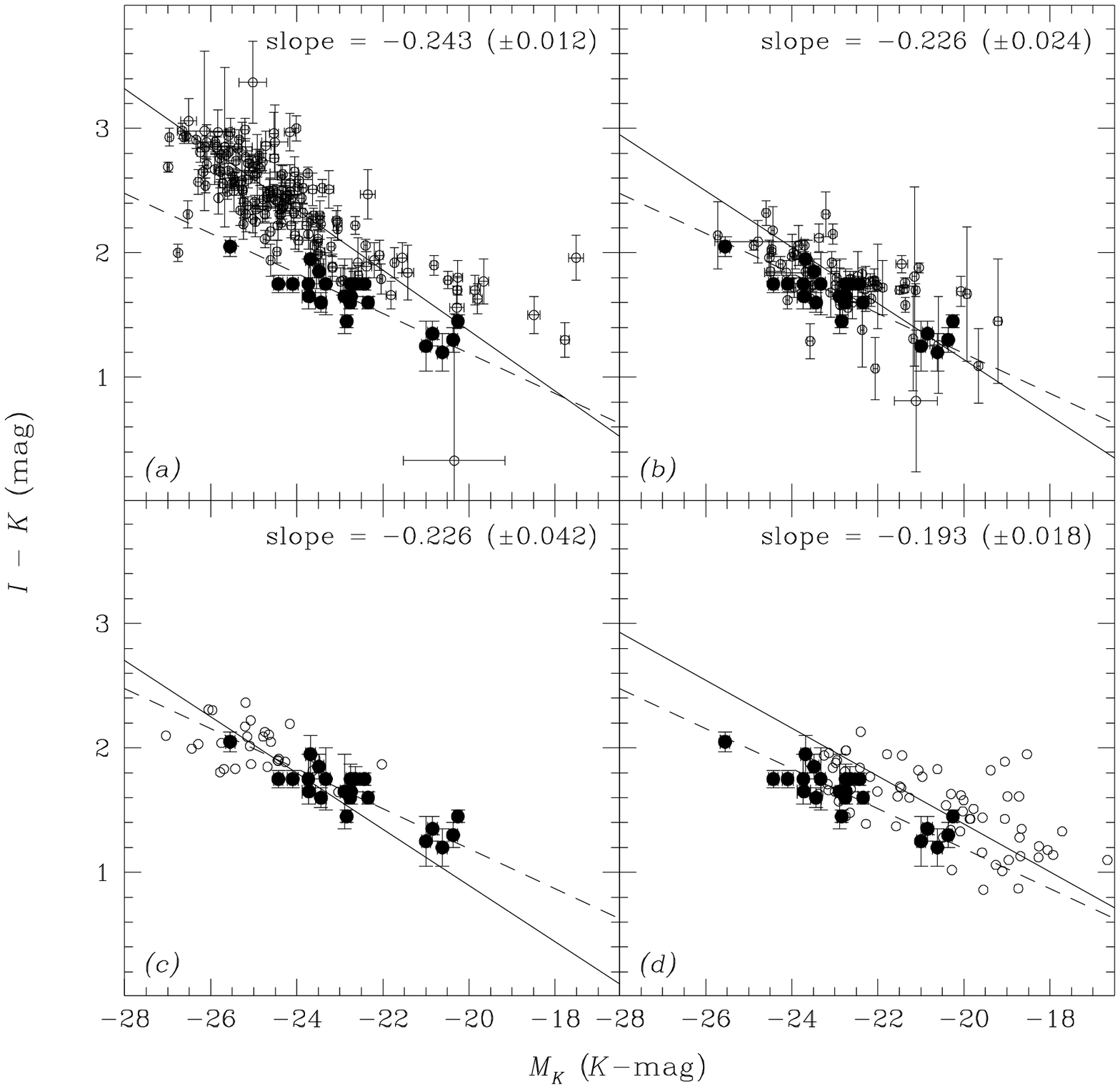}
\caption[]{\label{litcompare.fig}Comparison of our $M_K$ vs.  {\it
(I--K)} CM relation with previously published data.  The filled circles
show our sample galaxies; the open circles represent previously
published data.  The best-fitting NIR CM relations for the literature
samples are indicated by the solid lines; for comparison, the dashed
lines represent the mean CM relation derived from the data presented in
this paper.  {\it (a)} Bershady et al.  (1994), Bershady (1995); {\it
(b)} de Jong (1996b); {\it (c)} Andredakis et al.  (1995), Peletier \&
Balcells (1997); {\it (d)} Tully et al.  (1996).}
\end{figure*}

After scaling these and our observations to the same cosmology (using
$H_0 = 100$ km s$^{-1}$ Mpc$^{-1}$), we find that the most significant
deviations from our CM relation are exhibited by samples {\it (a)}
(Bershady et al.  1994, Bershady 1995), and {\it (d)} (Tully et al. 
1996). 

In either of these samples the reddening due to dust is unknown. 
Bershady et al.  (1994) and Bershady (1995) corrected their integrated
{\it I} and {\it K}-band colours neither for Galactic nor for internal
extinction in their sample galaxies.  Any comparison between our
dust-free and their integrated colours will therefore show an additional
discrepancy due to extinction effects.  Moreover, their redshift
distribution is different from those of the other samples, so that we
also have to take into account possible evolutionary effects when
comparing these relations. 

Tully et al.  (1996) corrected their photometry for both Galactic
foreground and internal extinction, using Tully \& Fouqu\'e's (1985)
inclination corrections.  Although this method is to first order useful
for low and intermediate inclinations, there are severe problems for
highly-inclined galaxies.  Moreover, for the analysis of a combined
sample of spiral galaxies, extinction effects should be treated
uniformly.  In addition, Tully et al.'s (1996) sample was drawn from a
cluster population, whereas the other samples consist (mainly) of field
galaxies.  Until it has been established unambiguously that
environmental effects do not play a role in the determination of the CM
relation for spiral galaxies, we cannot use a mixture of field and
cluster galaxy samples to base our conclusions on. Therefore, we will
consider field galaxy samples only in our analysis of the NIR CM
relation in this paper. 

Sample {\it (b)} (de Jong 1996b) consists of disc-dominated (nearly)
face-on galaxies.  The face-on orientation ensures that the effects of
internal extinction on the integrated {\it I--K} colours are small (but
not necessarily negligible), as opposed to more inclined galaxies, since
less light is obscured by the dust lane, because the light dominating
these colours is emitted by the old-disc stellar population located in
front of the dust component (which is concentrated in the galactic plane
region).  De Jong (1996a) argues that the colours and colour gradients
observed in his sample of face-on galaxies are determined by intrinsic
physical processes in the discs rather than by extinction effects. 
However, it is likely that the larger scatter in his CM relation
compared to ours is at least partly due to extinction effects affecting
his integrated colours (which are likely greater for these colours than
for our ``dust-free' colours, due to line-of-sight integration effects),
in addition to the effects of colour gradients across his face-on discs,
which vary as a function of galaxy type. 

Finally, sample {\it (c)} (Andredakis et al.  1995, Peletier \& Balcells
1997) consists of early-type disc galaxies (S0--Sbc), of which they
studied the disc and bulge components separately.  From the orientation
of the individual galaxies they obtained dust-free colours on the
non-obscured side of the galactic planes.  In addition, Peletier \&
Balcells (1997) assessed the importance of internal extinction by
studying the change in the position of the galactic centres as a
function of wavelength.  Such a shift occurs when a dust lane in front
of the true centre obscures more light on one side of the centre than on
the other, causing the observed centre to shift as a function of optical
depth, or passband.  Their centre shift is in general within the typical
errors, which provides additional support to the assumption that
extinction does not play a major role in their colour determinations. 

In Table \ref{slopes.tab} we compare the parameters of the NIR CM
relation derived from our sample of 22 edge-on disc galaxies with both
the CM parameters derived from the comparison samples in Fig. 
\ref{litcompare.fig}, and those obtained from supplementing our data
with samples {\it (b)} and {\it (c)}, the ``composite'' CM relation, for
which we are fairly confident that extinction effects are small. 

One might wonder to what extent the slope and zero point of our CM
relation are determined by the brightest, outlying galaxy.  We
redetermined these parameters for our sample, excluding this brightest
member, and found essentially identical results (see Table
\ref{slopes.tab}, for the sample ``without brightest'').  This was of
course to be expected, in view of the good agreement between our data on
the one hand, and the similar samples used to construct the composite CM
relation on the other. 

{
\begin{table}
\caption[ ]{\label{slopes.tab}{\bf Comparison of the
NIR CM parameters:}\newline
$(I-K) = a M_K + b$
\newline Columns: (1) Sample; (2) and (3) CM slope, {\it a}, and its
error; (4) and (5) Intercept, {\it b}, and its error; (6) Scatter in the
CM relation.}

\begin{center}
\begin{tabular}{lccccc}
\hline
\multicolumn{1}{c}{Sample} & {\it a} & $\pm$ & {\it b} & $\pm$ & $\sigma$ (mag) \\
\multicolumn{1}{c}{(1)}    & (2) & (3) & (4) & (5) & (6) \\
\\
{\it (a)} Bershady  & --0.243 & 0.013 & --3.48 & 0.30 & 0.256 \\
{\it (b)} de Jong   & --0.226 & 0.024 & --3.38 & 0.55 & 0.222 \\
{\it (c)} Peletier  & --0.226 & 0.042 & --3.63 & 1.04 & 0.173 \\
{\it (d)} Tully {\sl (all)} & --0.193 & 0.018 & --2.46 & 0.37 & 0.242 \\
{\it (d)} Tully {\sl ($T\ge4$)} & --0.199 & 0.025 & --2.67 & 0.51 & 0.256 \\
our                 & --0.156 & 0.021 & --1.99 & 0.48 & 0.133 \\
without brightest   & --0.157 & 0.024 & --1.99 & 0.55 & 0.137 \\
composite           & --0.189 & 0.013 & --2.59 & 0.31 & 0.205 \\
\hline
\end{tabular}
\end{center}
\end{table}
}

The dashed line in Fig.  \ref{composite.fig} shows the composite NIR CM
relation.  The best-fitting CM relation derived from our data alone is
shown by the dotted line.  The close agreement between both CM relations
supports our assumption that we can indeed compare our observations to
samples {\it (b)} and {\it (c)}.  In fact, the good agreement between
our sample and samples {\it (b)} and {\it (c)} leads us to note that
{\em the NIR CM relation for the old-disc population of spiral galaxies
in the field appears to be universal} (see Peletier \& de Grijs 1998).

\begin{figure*}
\vspace*{-9.5cm}
\psfig{figure=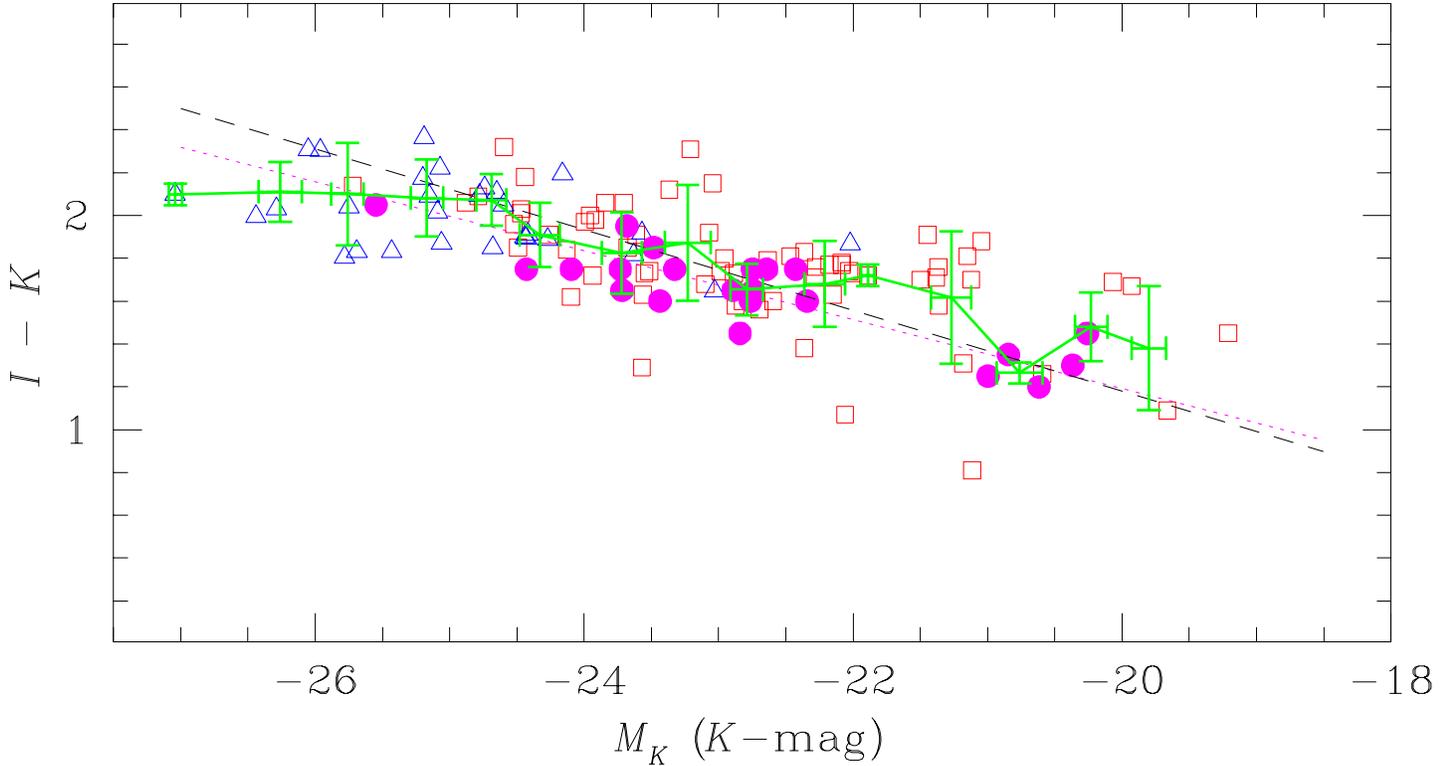}
\caption[]{\label{composite.fig}Composite CM relation, derived from
samples {\it (b)} and {\it (c)} and supplemented with the observations
presented in this paper.  Our observations are shown as filled circles;
the open squares correspond to sample {\it (b)}, the open triangles to
sample {\it (c)}.  The best-fitting composite CM relation is indicated
by the dashed line; its parameters are listed in Table \ref{slopes.tab}. 
The NIR CM relation derived from the observations presented in this
paper is indicated by the dotted line.  The solid line was obtained by
binning the data points in $M_K$ bins of 0.5 mag; the error bars
indicate the dispersion in the data points in each bin.} 
\end{figure*}

In Table \ref{slopes.tab} we also give the standard deviations of the
observed data points with respect to the best-fitting linear CM
relation, which are indicative of the scatter in these CM relations. 
The best-fitting relations were determined by means of a bivariate
minimization algorithm.  It is immediately clear that the scatter in the
CM relation determined from our dust-free data points is of order 40\%
smaller than for both samples {\it (b)} and {\it (c)}. 

Only if we can put firm constraints on the scatter in the CM relation
for field spiral galaxies, it may be useful as a distance indicator. 
Therefore, we compared the scatter in the CM relation that we obtained
from our dust-free disc colours to that from previous studies. 

From a detailed photometric study of Virgo and Coma cluster members,
Bower et al.  (1992b) found that for elliptical galaxies the scatter in
the CM relation is dominated by the observational errors.  The inclusion
of S0 and early-type spiral galaxies into their sample increases the
observed scatter in the optical-NIR ({\it V--K} ) CM relation.  To be
able to compare our results to those of Bower et al.  (1992b), we
applied the least-squares fitting technique we used to their data, and
found close matches between our and their values for the r.m.s. 
scatter.  Therefore, we can directly compare the scatter estimates
obtained from the samples listed in Table \ref{slopes.tab} to Bower et
al.'s (1992b) results. 

Although it is obvious that the scatter in samples containing spiral
galaxies is significantly larger than in samples containing only E/S0
galaxies (see also Visvanathan \& Sandage 1977, Peletier \& de Grijs
1998), we have shown that by avoiding the disturbing effects of the
in-plane dust lane the scatter in the spiral galaxy NIR CM relation can
be reduced significantly (e.g., Fig.  \ref{observations.fig}). 

Therefore, {\em a NIR CM relation for spiral galaxies based on
dust-free colours may in principle be useful as a diagnostic to estimate
distances.} It provides independent distance estimates based on
observational parameters. 

\section{Implications}

\subsection{The shape of the CM relation}

Lasker (1970) was the first to notice that the CM relation seems to
flatten towards the brighter galaxies in his sample, although in Fig. 
14 of de Vaucouleurs (1961) one can already see this effect.  Frogel et
al.  (1978) and Tully et al.  (1982) also note that the {\it V--K} colour
of their brightest sample galaxies is relatively independent of
luminosity compared to the fainter groups. 

With this relative flattening in mind, a linear fit to the CM relation
may not be the best representation of the correlation.  However,
flattening effects are second-order effects; to first order linear fits
should be sufficient for the determination of distance moduli. 

To investigate a possible flattening trend in our data we computed the
mean {\it I--K} colours in absolute magnitude bins of 0.5 mag, of which
the result is shown as the solid line in Fig.  \ref{composite.fig}; the
error bars indicate the dispersion in the data points in each bin. 
Although we are dealing with relatively few data points, a flattening is
indeed appreciated at the bright end of the relation.  This means that
the old stellar populations in the brightest -- or largest -- disc
galaxies have a roughly constant intrinsic {\it I--K} colour.  Thus, for
spiral galaxies with $M_K < -25.5$ the CM relation cannot be used as a
diagnostic tool for distance determinations.  At the faint end too few
data points are available to draw firm conclusions about the shape of
the CM relation, but it seems likely that the correlation is maintained
down to (field) spiral galaxies as faint as $M_K \approx -20$. 

\subsection{Applications of a tight CM relation}
\label{scatter.sect}

As we argued in Sect.  \ref{composite.sect}, a NIR CM relation for
spiral galaxies based on dust-free colours may in principle be useful as
a diagnostic to estimate distances: it provides independent distance
estimates based on observational parameters. 

For distance determinations to individual spiral galaxies, we have to
take into account the scatter in and the shallow slope of the CM
relation, as well as the observational errors, which will result in
relative distances accurate to $\sim 35$\% for the observations
presented in this paper.  This accuracy estimate is based on the
observed dispersion (i.e., the combination of observational and
intrinsic scatter) in the NIR CM relation {\it in the absolute K-band
magnitudes} ($\sim 0.7$ mag; i.e.  the dispersion along the horizontal
axis in Fig.  \ref{observations.fig}a).  For a ``typical'' sample
galaxy, the observational dispersion in the absolute magnitude
determinations (taking into account the uncertainties in the apparent
magnitude and the correction for the motion with respect to the Local
Group) is $\sim 0.45$ mag, which means that the intrinsic scatter in the
NIR CM relation among galaxies is $\sim 0.5$ mag for the absolute {\it
K}-band magnitude determinations.  Although it is likely that part of
this intrinsic scatter is caused by the peculiar velocities of our
sample galaxies with respect to the Hubble-flow movement of the centroid
of the Local Group, small-number statistics prevent us from reaching
conclusive evidence for this: peculiar motions are available for only 15
of our 22 sample galaxies (Table \ref{corrections.tab}; Mathewson et al. 
1992) For instance, correcting the absolute magnitudes for the effects
of the peculiar velocities results in a dispersion in $M_K$ of 0.994
mag, compared to 0.640 mag before correction of these 15 sample
galaxies.  The dispersion in $(I-K)$ increases from 0.102 to 0.161 mag,
based on bivariate fits. 

The observational uncertainties in the apparent {\it K}-band magnitudes
discussed in this paper are largely due to the relatively shallow
(``snap-shot'') nature of the observations, as well as to the rapidly
varying sky background in the {\it K} band (see de Grijs [1997], and de
Grijs et al.  [1997] for a discussion of the relevant photometric
accuracy achieved in this project).  This rapid variability requires
that short ($< 60$s) object and sky observations, of equal integration
times, are taken alternately, in order to correct for these background
variability effects.  Obviously, a greater number of such observations
will smooth out any remaining effects due to the varying background on
the one hand, and increase the signal-to-noise (S/N) ratio in the
galactic outskirts on the other, thereby improving the photometric
accuracy of the integrated {\it K}-band magnitudes significantly.  In
addition, to unambiguously establish a uniform spiral galaxy CM
relation, the distances to the sample galaxies need to be known to high
accuracy.  For our present sample, uncertainties in the absolute {\it
K}-band magnitudes can be reduced by the determination of distances that
are more accurate than those used in this paper, which are based on
recessional velocities.

Therefore, by obtaining high-quality observations of and improved
distances to moderately or highly inclined galaxies, the accuracy of
this method can be enhanced by reducing the observational scatter in the
apparent {\it K}-band magnitudes to a lower limit of $\sim 0.02$ mag. 
The accuracy will thus be limited by the intrinsic dispersion in the NIR
CM relation in the absolute {\it K}-band magnitudes ($\sim 0.5$ mag). 
Because of this intrinsic scatter, the maximum accuracy of distance
determinations to spiral galaxies that can be reached using this
relation will therefore be $\sim 25$\%.  Note that this accuracy
approaches the 20\% accuracy of Bower et al.  (1992b) for early-type
galaxies, for which the intrinsic scatter in the CM relation is
significantly smaller. 

The intrinsic dispersion in absolute {\it K}-band magnitudes in the CM
relation for field spiral galaxies is most likely caused by the
non-negligible effects of different degrees of active star formation,
star formation histories, ages, metallicities, and extinction, even at
the {\it z} heights at which we determined our ``dust-free'' colours,
and perhaps non-negligible vertical colour gradients in the discs of our
sample galaxies (Peletier \& de Grijs 1998; de Grijs \& Peletier, in
preparation).  In view of the expected variation of all these
parameters, it is even more surprising to find a very tightly
constrained NIR CM relation for field spiral galaxies. 

\subsubsection{How {\sl Hubble Space Telescope} observations can
contribute to the calibration}\label{hst.sect}

Since the galaxies need to be spatially resolved, high-resolution
observations done with the {\sl Hubble Space Telescope (HST)} can
provide a powerful tool to minimize the observational scatter and extend
the useful distance range.  Such observations may therefore provide the
means to calibrate the relation. We will demonstrate this in the
following exercise:

\begin{itemize}

\item {\it Model galaxy.} We assume that the intrinsic major-to-minor
axis diameter of a ``typical'' spiral galaxy disc is 9 : 1 (e.g.,
Guthrie 1992), and the scale height ratio of the dust to the stellar
disc is 1 : 2 (following the standard models in Huizinga 1994; see also
Xilouris et al.  [1999] for case studies), whereas the dust disc is
embedded in the stellar disc according to the so-called ``triplex
model'' (e.g., Disney, Davies \& Phillipps 1989).  Furthermore, either
disc is assumed to be of exponential form, both radially and vertically. 

\item {\it Inclination restrictions.} The requirement that we need to be
able to control -- to some extent -- the effects of extinction restricts
the useful inclination range to inclinations between $\sim 80^\circ$ and
$90^\circ$.  The lower inclination limit is set by the fact that at $i
\simeq 80^\circ$ the dust lane will be observed at the outer rim of the
projected galactic disc nearest to the observer.  This implies that,
from this inclination upwards, the upper stellar layer at the far edge
of the disc will not significantly be affected by the effects of
extinction.  For higher inclinations, an increasingly large dust-free
area will be available for colour measurements, divided in two parts by
the high-extinction in-plane dust lane. 

\item {\it Size restrictions.} For an unambiguous determination of these
dust-free colours, one will ideally need a dust-free area of at least
three resolution elements containing galactic emission brighter than the
limiting magnitude or S/N ratio.  With the superb angular resolution of
the {\sl HST}, we will be able to reduce the minimum projected sizes of
the useful candidate galaxies considerably.  In the NIR, the most
suitable {\sl HST} instrument is the {\sl Near-Infrared Camera and
Multi-Object Spectrometer (NICMOS)}, which has a pixel size of $0.''075$
(camera 2) and a FWHM for point sources of $\sim 1.3$ pixels $(0.''10)$
at 1.1$\mu$m to 2.5 pixels $(0.''19)$ at 2.2$\mu$m (Krist \& Hook 1997). 
The inclusion of a red optical passband (e.g., {\it I} ) requires
additional imaging with the {\sl Wide Field Planetary Camera 2 (WFPC2)};
the pixel scale of the {\sl Planetary Camera} is $0.''0455$, with a
point source FWHM of about 2.5 pixels ($\sim 0.''11$, depending on
wavelength).  Since the {\sl WFPC2} characteristics fall inside the
parameter range for {\sl NICMOS} camera 2, we will restrict ourselves to
the possibilities provided by the use of {\sl HST NICMOS} observations
in the remainder of this exercise. 

\end{itemize}

Along the galaxies' minor axes, the useful area for colour determinations
is, for all inclinations ($80^\circ \le i \le 90^\circ$) and assuming a
minimum dust-free extent of 3 resolution elements on the side nearest to
the observer, $\sim 0.''29$ and $\sim 0.''56$ for the {\it J} and {\it
K} bands, respectively.  Under the additional assumptions outlined at
the beginning of this paragraph, and by using Hubble's (1926) relation
between a galaxy's inclination and its axis ratio, the minimum projected
size of a candidate galaxy in the {\it J} band is $1.''6 \times 8.''5$
to $1.''2 \times 10.''5$ for inclinations of $80^\circ$ to $90^\circ$;
in the {\it K} band the different FWHM converts to projected sizes of
$3.''0 \times 16.''4$ to $2.''3 \times 20.''2$ for the same inclination
range.  We are thus limited by the sizes required by the {\it K}-band
imaging. 

Because of the high resolution of the {\sl HST} one will be able to
resolve a galaxy of a given size (determined by the limiting magnitude
or S/N ratio of the observations) out to larger distances than when
observed with (seeing-limited) ground-based telescopes.  To get an
indication of the usefulness of the NIR CM relation as a function of
distance we obtained {\sl HST} archive observations of NGC 891.  This
edge-on Sbc galaxy is comparable in size and appearance to the Galaxy. 
We used observations in the optical F606W passband (160s) and the NIR
F160W filter (192s). 

In the (wide) {\it V}-band observations, the maximum useful dust-free
region on the minor axis, where the galactic {\it V}-band surface
brightness $\Sigma_V \le 25$ mag arcsec$^{-2}$, is $(33.''0 \pm 0.''5)$,
which corresponds to $(1.50 \pm 0.02)$ kpc at its distance of 9.5 Mpc
(van der Kruit \& Searle 1981, see also Keppel et al.  1991).  The
minimum dust-free region on a galaxy's minor axis containing three
independent resolution elements is $0.''56$, which is required by the
need for {\sl NICMOS K}-band imaging, as explained above. 

From this argument it follows immediately that we can use galaxies of
similar size as NGC 891 for the construction of a NIR CM relation up to
distances of $(560 \pm 5)$ Mpc.  For comparison, ground-based
observations can be used for similar galaxies out to distances of only
$\sim 210$ Mpc, under optimal seeing conditions (FWHM $\sim 0.''5$).  In
Fig.  \ref{distances.fig} we show the difference between the results
obtained using {\sl HST} observations and ground-based observations
taken under such optimal seeing conditions for edge-on galaxies like
those presented in this paper.  To construct this figure, we have
adopted the assumptions about the galactic structure outlined above, and
a major axis diameter of $D_{25}^B = 13.'5$ for NGC 891 (RC3),
corresponding to a linear size of 37.3 kpc. 

\begin{figure}
\psfig{figure=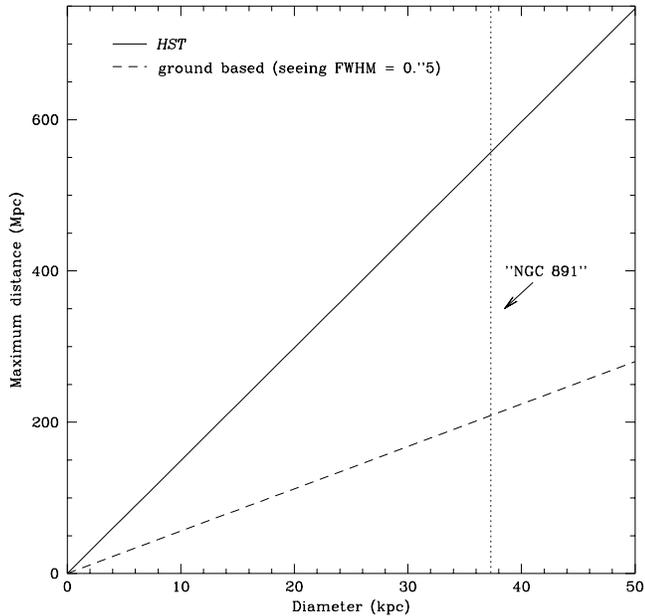,width=9cm}
\caption[]{\label{distances.fig}Comparison between the maximum distances
out to which the NIR CM relation can be used successfully to obtain
distance estimates for {\sl HST} and ground-based observations under
optimal seeing conditions.  The power of space-based observations for
distance determinations using this relation is clear.}
\end{figure}

From a comparison between the optical and the NIR {\sl HST} observations
of NGC 891, with integration times of the same order, it is clear that
the NIR observations do not reach the galactic outskirts at similar S/N
ratios as in the optical.  Therefore, if we base our conclusions on this
particular case, the successful use of the NIR CM relation for distance
estimates depends on the faintest surface brightness levels that can be
achieved in the NIR.  However, this does not need to be the case for all
galaxies, as it depends on parameters, such as intrinsic colours, sky
background (which is supposed to be very low for space-based NIR
observations), etc. 

In summary, high-resolution observations done with the {\sl HST} can
significantly increase the number of useful galaxies and hence provide a
powerful tool to calibrate the CM relation for spiral galaxies, reduce
the observational scatter, and extend the useful distance range by more
than a factor of 2 compared to ground-based observations.  The useful
projected galaxy sizes are limited by the {\it K}-band resolution, which
translates to minimum sizes from $3.''0 \times 16.''4$ to $2.''3 \times
20.''2$ for inclinations of $80^\circ$ to $90^\circ$.  Obviously, {\it
I} (or {\it J}) and {\it K}-band observations of galaxy fields from the
{\sl HST} data archive (e.g., the {\sl Hubble Deep Fields}, among
others) would be well suited to undertake such a programme. 

\section{The NIR Tully-Fisher relation}

For the majority of our sample galaxies a homogeneous data set
containing both the maximum (optical) rotational velocities and total
(apparent) {\it I}-band magnitudes was provided by Mathewson et al. 
(1992) and Mathewson \& Ford (1996).  In de Grijs (1997, 1998), we
compared our {\it I}-band photometry to that of Mathewson et al.  (1992)
and Mathewson \& Ford (1996).  From the detailed comparison of our
photometry to theirs it was shown that we can accurately reproduce their
results ($\langle m_{I,{\rm our}} - m_{I,{\rm Mathewson}}\rangle = -0.07
\pm 0.13$; de Grijs 1997, 1998). 

The {\it I} and {\it K}-band TFRs derived from our absolute magnitudes,
and supplemented with Mathewson's rotational velocity data, are shown in
Fig.  \ref{tf.fig}.  We did not correct our integrated magnitudes for
the effects of internal interstellar extinction; inclination corrections
that aim to correct internal extinction are generally not applicable to
highly inclined galaxies.  Moreover, the amount of extinction in a
galaxy varies as a function of galaxy type, as well as among galaxies of
the same type (see, e.g., de Grijs et al.  1997, de Grijs 1997, 1998,
and references therein).  It is therefore likely that the scatter in the
{\it I} and {\it K}-band TFRs derived from our observations is at least
partly caused by internal extinction.  However, note that errors in the
TF distances are generally dominated by uncertainties in the
inclinations of the sample galaxies. 

\begin{figure}
\psfig{figure=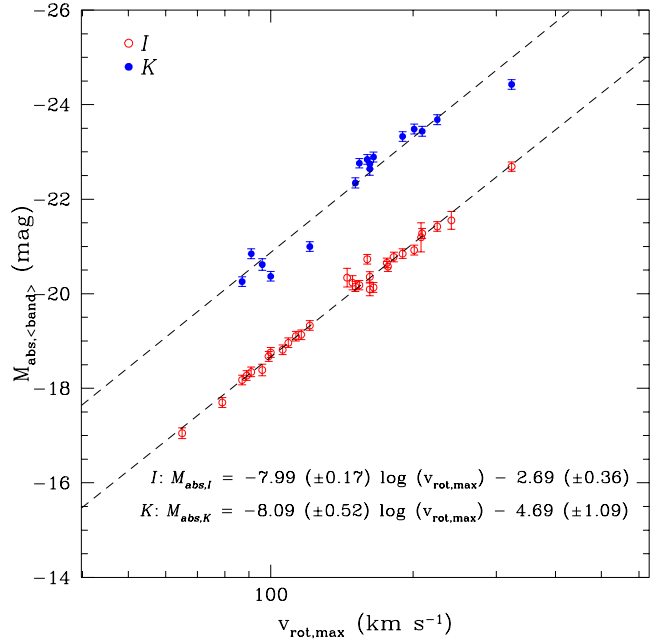,width=9cm}
\caption[]{\label{tf.fig}{\it I} and {\it K}-band TFRs derived from our
observations (no internal extinction correction was applied).}
\end{figure}

\subsection{The scatter in the Tully-Fisher relation}

A good measure of the accuracy of the TFR can be obtained by assessing
the scatter in the relation.  In the {\it I} and {\it K} bands, the TFR
r.m.s.  scatter is 0.145 and 0.296 mag, respectively, based on bivariate
fits (e.g., Giovanelli et al.  1997, Verheijen 1998).  The scatter in
the {\it K}-band TFR is significantly larger than that in the {\it I}
band.  This is probably caused by the fact that both the S/N ratio in
our $K'$-band images is significantly smaller and the effects of the
varying sky background are much greater in $K'$ than in the {\it I}-band
observations, causing the larger uncertainties, and thus observational
scatter, in the {\it K}-band magnitudes.  We are confident that the
scatter in the {\it K} band TFR due to internal extinction is very small
in comparison, since the {\it K}-band vertical surface brightness
profiles exhibit hardly any signatures of either a regular dust lane or
a patchy dust distribution, even in those galaxies showing pronounced
dust lanes in the optical passbands (de Grijs et al.  1997). 

The r.m.s.  scatter in our {\it I}-band TFR is small compared to that
found in previous studies (for a discussion see Bernstein et al.  1994,
Giovanelli et al.  1997): only Bernstein et al.  (1994) reported a
significantly smaller scatter, of $\sigma = 0.10$ mag, based on a sample
of 22 galaxies.  Giovanelli et al.  (1997) argue that TF fits with
scatter smaller than 0.25 mag are likely to be statistical accidents,
which can occur when galaxy samples are small.  They base this argument
on the detailed study of the TFR for a large number of galaxies in 24
clusters.  Giovanelli et al.  (1997) determined the scatter in the {\it
I}-band TFR for the individual clusters; their observational scatter
varies between 0.12 and 0.36 mag (based on bivariate fitting routines). 
With this statistically significant result in mind, we believe that the
low scatter found in this paper is indeed a statistical accident. 
Additional support for this assumption is lent by the fact that the
scatter in the CM relation increases (in either of the quantities) if we
correct for the effects of peculiar motions, as shown in Sect. 
\ref{scatter.sect}. 

To our knowledge, studies of the TFR in the NIR {\it K} (or $K'$) band
are scarcely available to date.  From Fig.  10 of Tully \& Verheijen
(1997), we derive a scatter of 0.531 mag in the $K'$-band TFR of Ursa
Major cluster galaxies, based on a bivariate fit; for the low and high
surface brightness galaxies in this figure, we derive a scatter of 0.654
and 0.363 mag, respectively.  In addition, in a detailed study of a
volume-limited complete sample of $\sim 40$ Ursa Major cluster galaxies
Verheijen (1997) finds that the smallest observed TFR scatter in his
galaxy sample is 0.29 mag in $K'$, consistent with no intrinsic scatter. 

Giovanelli et al.  (1997) applied a type-dependent correction for
internal extinction to their observations, which varies between 0.50 and
1.00 mag in the {\it I} band, corresponding to 0.12 -- 0.23 mag in the
{\it K} band (assuming an $R_V = 3.1$ extinction law, e.g., Schlegel et
al.  1998).  If we apply a similar extinction correction to our data, we
find a scatter in the {\it I}-band TFR of 0.225 mag (as opposed to 0.145
mag for the uncorrected measurements), and a scatter of 0.316 mag in the
{\it K}-band TFR (compared to 0.296 mag before extinction correction),
which is entirely consistent with Giovanelli et al.'s (1997) results. 

From their detailed analysis of the scatter in the TFR, Giovanelli et
al.  (1997) concluded that the average total scatter in their
measurements of a statistically significant number of galaxies in 24
clusters is $\sim 0.35$ mag.  If we assume that this is a typical value
for the scatter in any well-sampled TFR, the accuracy of the TF distance
to a ``typical'' galaxy in our sample would amount to $\sim 15$\%. 

When comparing our TF results with the accuracy of distance
determinations using the NIR CM relation, we note that, due to the
shallower slope of the CM relation, the effects of uncertainties in the
distance estimates are relatively more important for the calibration of
the spiral galaxy CM relation than for the TFR.  On the other hand, one
of the main advantages of the use of the NIR CM relation is that {\em we
only need photometric data} to obtain distance estimates, whereas use of
the TFR requires additional kinematic data.

\subsection{Distance estimates: Tully-Fisher vs. Colour-Magnitude}  

In Sect.  \ref{hst.sect} we estimated that the NIR CM relation can be
used to obtain distances to galaxies similar to the Galaxy or NGC 891 up
to $(560 \pm 5)$ Mpc, if {\sl HST}-based photometry and resolution can
be obtained.  Following similar lines of argument, in this section we
will estimate the corresponding maximum TF distance to a Galaxy-type
system that can still be determined unambiguously.  To do so, we will
restrict ourselves to ground-based rotation curve measurements, since
they are most readily available in the literature, and most easily
obtained. 

Based on a statistical analysis of a spiral galaxy sample designed for
TF applications, Courteau (1997) has shown that the best measure of TF
velocity is given at the location of peak rotational velocity of a pure
exponential disc.  Alternatively, one can use the 20\% width of the
H{\sc i} velocity profile.  Although this latter method does not make
any {\it a priori} assumptions about the luminosity profile or shape of
the rotation curve, its accuracy is not optimal (Courteau 1997). 

If we restrict ourselves to optical observations, one needs at least 5
resolution elements with reliable S/N ratios to obtain an accurate
measure of the rotation curve.  Under optimum seeing conditions (FWHM
$\sim 0.''5$), this corresponds to a minimum angular diameter of a
galactic disc of $\sim 2.''5$.  A Galaxy-type system like NGC 891, with
a linear diameter of 37.3 kpc (Sect.  \ref{hst.sect}), would therefore
be useful for distance estimates up to $\sim 3.1$ Gpc ($\sim 5.5$ times
as far as the maximum CM distances).  Obviously, the use of H{\sc i}
observations will increase this distance estimate significantly. 

The main disadvantage of the use of the TFR versus the CM relation for
distance estimates to relatively nearby, {\it highly inclined} galaxies
is that the conversion of apparent to face-on corrected magnitudes is
controversial and model dependent.

\section{Summary and Conclusions}

In this paper we have looked at two methods of distance determination to
spiral galaxies in the field using optical/NIR observations, the {\it
(I--K)} vs.  $M_K$ colour -- absolute magnitude relation and the {\it I}
and {\it K}-band Tully-Fisher relation.  We have used ``dust-free''
colours to achieve greater accuracy for distance determinations using
the CM relation compared to the integrated galaxy colours that are
generally used for this exercise.  Our main conclusions are the
following:

\begin{itemize}

\item Our data, supplemented with observations taken from the literature
form a well-constrained composite spiral galaxy CM relation; it appears
that the NIR CM relation for the old-disc population of spiral galaxies
in the field is universal. 

\item By avoiding the disturbing effects of the in-plane dust lane the
observational scatter in the CM relation can be reduced significantly. 
Therefore, the NIR CM relation for field spiral galaxies, based on
dust-free colours, may in principle be useful as a diagnostic tool to
estimate distances with an accuracy of $\sim 25$\%.  This accuracy is
limited by the intrinsic dispersion of the NIR CM relation in the {\it
K}-band absolute magnitudes ($\sim 0.5$ mag). 

\item High-resolution observations done with the {\sl Hubble Space
Telescope} can provide a powerful tool to reduce the observational
scatter, calibrate the relation, and extend the useful distance range
(by more than a factor of 2 compared to ground-based observations).  The
useful projected galaxy sizes are limited by the {\sl NICMOS} {\it
K}-band resolution, which translates to minimum sizes from $3.''0 \times
16.''4$ to $2.''3 \times 20.''2$ (at the limiting magnitude or S/N
ratio) for inclinations $80^\circ \le i \le 90^\circ$. 

\item We found a scatter in the {\it I} and {\it K}-band TFRs of 0.145
and 0.296 mag, respectively.  Although, to date, these values are among
the lowest found in these passbands, they are likely statistical
accidents, caused by our relatively small sample size (see Giovanelli et
al.  1997). 

\item Due to the shallower slope of the CM relation, the scatter is
relatively more important for the determination of distances via this
method than for distance determinations using the TFR: typically, TF
distances can be determined with an accuracy of $\sim 15$\% (and to
significantly greater distances), as opposed to the $\sim 25$\% accuracy
achieved using the NIR CM relation.  However, one of the main advantages
of the use of the NIR CM relation is that {\em we only need photometric
data} to obtain distance estimates, whereas use of the TFR requires
additional kinematic data. 

\item Although a linear fit is a good first-order approximation to the
composite NIR CM relation, the old-disc populations of the brightest
disc galaxies ($M_K < -25.5$) have a roughly constant intrinsic {\it
I--K} colour.  Our observations at the faint end of the CM relation are
consistent with a linear correlation down to galaxies as faint as $M_K
\approx -20$. 

\end{itemize}

\section*{Acknowledgments} 
We acknowledge discussions with Piet van der Kruit and Rob Swaters. 
During part of this work, RdeG was supported by NASA grants NAG 5-3428
and NAG 5-6403.  We have made use of the new {\sl DIRBE/IRAS} dust maps
made available electronically by David Schlegel et al.

\end{document}